# iPaaS in Agriculture 4.0: An Industrial Case

Rafael Huang Cestari
Université de Pau et des Pays de l'Adour, E2S UPPA, LIUPPA, Anglet, France
rafacestari@poli.ufrj.br

Sebastien Ducos
Université de Pau et des Pays de l'Adour, E2S UPPA, LIUPPA, Anglet, France
s.ducos@maisadour.com

Ernesto Exposito
Université de Pau et des Pays de l'Adour, E2S UPPA, LIUPPA, Anglet, France
ernesto.exposito@univ-pau.fr

***Abstract*—Current automation approaches in the Industry 4.0 have generated increased interest in the utilization of Integration Platforms as a Service (iPaaS) cloud architectures in order to unify and synchronize several systems, applications, and services in order to build smart solutions for automated and adaptive industrial process management. Existing iPaaS solutions present several out-of-the-box connectors and automation engines for easier integration of customers' projects, but show issues regarding overall adaptation outside their scope, brand locking, and occasionally high prices. Moreover, existing platforms fail to respond adequately to the needs of deploying multiple decision models capable of offering automated or semi-automated management of processes, thanks to the integration of the large diversity of data and event sources as well as the different physical or logical action entities. With the popularization of open-source software and applications such as BPM Engines, Machine Learning libraries, and Integration suites and libraries, it is possible to develop a fully customizable and adaptable, open-source iPaaS that can be used both in and outside industrial applications. In this paper, we propose a generic iPaaS architecture implemented on the basis of several open source solutions boasting integration, interoperability, and automated decision-making capabilities in the domain of Agriculture 4.0. A proof-of-concept based on these solutions is presented, as well as a case study on MAÏSADOUR's grain storage process with a comparison with the currently human-operated tasks.**

## I. Introduction (*Heading 1*)

The needs for automation of agricultural activities has been an urgent necessity in the context of the new Industry 4.0 revolution. In order to keep up with the ever-growing global population and market demands, new technologies, methods, and techniques are constantly developed in order to maximize production efficiency. Various devices and systems, such as automated harvesters, environment monitors, and humidity and temperature controllers are utilized in several agricultural applications to great effect, assuring efficient cultivation, gathering and storage of products. However, when considering the expansion of these applications and the amount of data generated, the cohesion and cooperation between these systems can be improved and expanded, allowing for a greater scope of automation and logistics management. A promising solution is represented by the Integration Platform as a Service (iPaaS), which is a suite of cloud services that enables managing and integration flows by connecting a wide range of applications or data sources without manually installing or managing any hardware or middleware [1]. This allows for the creation of scalable, efficient, and cohesive systems designed for automated activities, data evaluation, and decision-making, which includes not only contextual devices, but also the resulting products and the involved human participants as well, following the principles of the Internet of Everything (IoE) concept [2]. It also guarantees the interoperability of present and future systems by guaranteeing smooth context adaptation and system evolution.

Beyond the scope of integration and automation, there is also the possibility of increasing process efficiency with the use of Machine Learning (ML) modules and methods, taking in and analyzing data in order to improve its decision-making capabilities, making it faster, more precise, cheaper, or more reliable, as it can reduce the amount of human interaction needed within the loop.

In this paper, we present our findings in the framework of an industrial project aimed at designing and prototyping an iPaaS solution well-suited for grain storage and processing capable of integrating heterogeneous actors such as data collectors, temperature and humidity controllers, in-site grain transport systems, transport personnel tracking, business process engines, databases, among others. The main goal of this solution is to reduce energy consumption, as well as reduce the need of interaction by human experts in the process. This solution has been designed in order to facilitate the integration of additional systems, making it possible for it to be used in applications other than those focused on agricultural activities, while also retaining the ease of intercommunication between systems and the ability to provide automated decision-making with Business Process Management (BPM) engines and Machine Learning (ML) modules.

This paper is organized as follows: In Section II, the state of the art is laid out, as well as the needs and goals of the expected IPaaS solution. In Section III, the proposed solution is characterized and defined. In Section IV, the implementation of the solution is explained and detailed. Finally, in Section V, results are presented and in Section VI, final conclusions and perspectives are made.

## II. Integration Platform as a Service

An iPaaS is a cloud-based solution for integration of different applications, services, private and public clouds, and systems without having to invest time and resources into installing, developing and maintaining integration software, middleware and hardware [3]. The iPaaS facilitates development of business systems, leveraging of data (especially large amounts), reducing time-to-value and installation costs, and requires less specialized skills to be deployed.

This paper focuses on an Industrial Case belonging to MAÏSADOUR, an international food cooperative that deals in agricultural, pork, fish, and poultry products, among others [4]. The case in question is the process of drying and storing grains that have been recently harvested.

It is required for the iPaaS solution to possess integration and interoperability features, such as connectors and messaging patterns, for the unison of the different actors in the process, a business process engine or management features for automated coordination of the process, and a decision engine or ML functionalities for automated decisions and improvement of the process, also reducing the need for human interaction. Additionally, the solution should, have a low or reasonable cost, allow for full control of its functions, and provide ease of usage and implementation. Several companies offer iPaaS solutions and services that include these features. A brief analysis of the most used and reviewed services was made utilizing information from Gartner provided from their Peer Review [5] website and "Magic Quadrant" 2019 report [6].

Microsoft Azure Integration Services is composed of several Azure services such as Azure Logic Apps, Azure API Management, and Azure Service Bus. Data integration and workflow automation are also offered separately. These services can make use of Enterprise Integration Patterns (EIP) [7] and several messaging patterns [8]. There are more than 360 different connectors available between these services [9]. It has innate integration with other solutions provided by Microsoft for these use cases, but has brand locking problems.

The Dell Boomi platform includes several functionalities, such as core application/data integration, API and Business-to-Business management, and workflow automation and application development. Different offerings are available, geared towards specific use cases and with a set amount of endpoints. Boomi has over 1500 unique connectors available [10] as well as different integration patterns [11]. Customer support is considered lacking, and there are missing features such as data virtualization and event stream analytics.

IBM Cloud Integration is composed of several services such as IBM API Connect, App Connect, MQ, Event Streams, IBM Aspera, Secure Gateway Service, IBM BPM and others. It boasts standard messaging functions [12] and over a hundred built-in connectors with the possibility of the customer developing custom ones [13]. It has a large scope of applications, being adaptable for various use cases when paired with other solutions provided by IBM. It also provides both free and paid versions. Implementation of this platform is considered difficult, and support provided has been criticized by customers.

Informatica Intelligent Cloud Services (IICS) is a microservices-based iPaaS that provides several fundamental functionalities, such as Business-to-Business integration, API management, and data integration. Several messaging patterns [14] and over 400 pre-built connectors are available as well as several custom ones [15]. It is considered to be more expensive than its competitors, and lacks certain features such as event processing.

As it can be observed from the information in the analysis and in Tables I, most of the solutions fulfill the criteria established for our case. They are widely accepted by several customers and provide automation and customer support, albeit said support is considered lacking. Integration with third-parties is also difficult. The use case-specific nature of some offerings also hampers adaptability. Finally, most of these solutions are entirely dependent on the development on the provider's side, reducing the level of control that the customer has.

TABLE I. SOLUTION COMPARISON (FUNCTIONAL FEATURES)

| Solution | Overview | Process & Decision Engine |
|---|---|---|
| Microsoft | Pub/Sub, Request-Reply, Queues, Point-to-Point, API/WebServices(WS) , Over 360 connectors, Paid, Includes Support, limited control | Yes/Yes |
| Dell | Point-to-Point, Pub/Sub, API/WS, Over 1500 connectors, Paid, Includes Support, limited control | Yes/No |
| IBM | Pub/Sub, Sync & Async, Over 100 connectors + custom , Paid & Free, Includes support, limited control | Yes/Yes |
| Informatica | Request-Reply, Pub/Sub, Queuing, One way, WS. Over 400 connectors + custom, Paid, Includes support, limited control | Yes/Yes |

Based on this study, it appears that the needs for a generic, open and extensible iPaaS platform have not yet been met. Such generic solution should allow users to have full access to all of the platform's functionalities, allowing its specialization based on new connectors, features and systems well-adapted to a large diversity of industrial contexts. Moreover, this also means that some of the ease-of-use advantages of proprietary solutions are also lost, which results in a steeper learning curve or higher entry-level knowledge, as well as longer development time required to fully cope with specific needs for industrial process management.

## III. PROPOSED SOLUTION

As stated previously, there are real needs to design and implement an open, generic and extensible iPaaS platform, based on the integration or development of a large diversity of software and services. In order for this proposition to be viable, it needs to address three key capabilities: Integration, allowing systems to communicate with each other, Planning and Execution, enabling automation in decision making and following a general business process model, and Monitoring and Analysis, evaluating received data and outcomes and proposing or implementing solutions to control the managed process automatically.

In order to design such an iPaaS solution, the ARCADIA approach was followed, which is a system engineering method based on the use of models, with a focus on the collaborative definition, evaluation and exploitation of its architecture [18]. Due to space limitations, the initial operational and system analysis phases will not be presented in this article. These phases have allowed us to identify the functionalities and the generic internal architecture of the IPaaS platform that is illustrated by the Logical Architecture diagram presented in Figure 1.

The iPaaS is composed of three main logical components: the integration module, the process manager, and the prediction module. Logical actors relevant to the architecture are the data collector, which can be also seen as the workspace or environment, the data storage, the boiler and sensors, and the grains. The integration module receives and sends information between the various components and actors involved in the process, guaranteeing that messages and data are formatted and relayed accordingly. The process manager controls the flow and decisions of the process, coordinating it. Finally, the prediction module, as the name implies, is responsible for the prediction of relevant parameters from data acquired from by the data collectors.

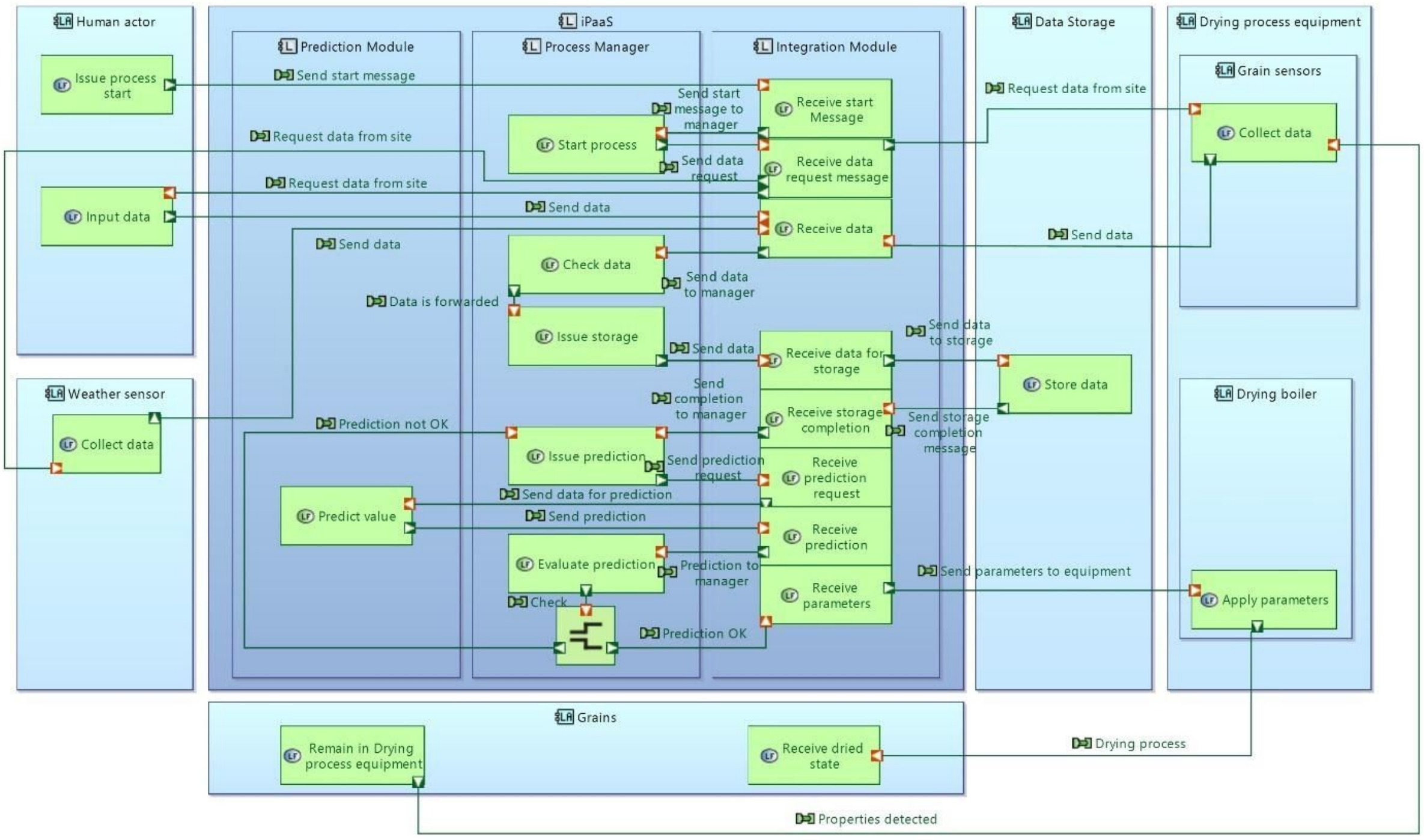

Fig. 1. Logical Architecture of the solution

Once the logical architecture phase was achieved, the identification of concrete components and subsystems able to satisfy the expected functional and structural requirements of the platforms was carried out. During the Physical architecture phase a set of components were identified as satisfying the expected requirements, in particular: the Java-based Camel framework (v2.25.0) for building system integrations [19], Camunda BPM (v7.13), which is a Business Process Management Notation (BPMN) Engine and editor capable of automation [20], and Apache Kafka (v2.12 - 2.4.1), which is a distributed streaming platform facilitating real-time data sources integration [21]. The Physical Architecture can be seen on Figure 2. Camel fulfills the role of integration, allowing for easier communication between different applications and services, and guarantees interoperability by having several different components (connectors) that allow for easy communication between systems, and constant updates and additions to said components in order to comply with newer standards. Camunda is responsible for automating the process through BPMN diagrams, as well as laying out the whole process with a comprehensive diagram, which is interchangeable with other BPMN diagram editors and engines. Finally, Kafka deals with any additional communication that has to be done and cannot be easily solved with a Camel application. Comparisons were made between the chosen components and possible alternatives as to prove that these choices were better suited for the project. However, due to this paper's limits, these comparisons will be left out.

For automated data processing and decision-making, a ML module was written in Python utilizing Tensorflow (v1.15.2), an open-source machine-learning platform [22], implementing Adaptive-Network-Based Fuzzy Inference System (ANFIS) [23]. In addition to these components of the iPaaS, a CouchDB database was also utilized for the storage of necessary data, which will be discussed later on.

In regard to the scalability of the iPaaS, it was decided, after a review of diverse cloud technologies, which is not discussed here due to this article's limits, to use Docker and Kubernetes as the main components to manage scalability and elasticity requirements. In the next section, a proof-of-concept focused on the integration, interoperability, and decision management and learning capabilities will be presented.

## IV. Implementation

One iPaaS prototype implementation has been built in order to manage agriculture processes while minimizing energy consumption. This prototype uses Camel as its main integrator solution. Java components are built using the Camel library in order to make communication between existing and future components easier. The prototype is based on the integration of CouchDB, Camunda BPM, Apache Kafka, and information from Openweather's API through appropriate connectors (such as ones dedicated to CouchDB and Kafka) or through HTTP REST API calls (which is the case for Camunda and the weather API).

A generic agriculture process, managed by Camunda from within a WildFly server, that represents the grain storage case has been implemented. As it can be seen in Figure 3, the process diagram consists of several tasks and gateways that must be completed and satisfied in order for the process to move forward. The first tasks are the data collection tasks that gather the weather, human and IoT data relevant to the drying and storage process. These must be completed in parallel to each other. Following that, the data is saved in CouchDB and then run by the ML module to acquire the parameters and setpoints necessary for the drying process (temperature, extraction time, humidity level, humidity goal) while minimizing energy consumption. This communication is done with Apache Kafka topics as intermediaries between the Camel application and the ML module. Once the predicted parameters and outcomes are acquired by the Camel application, they are then returned to the Camunda engine, which then decides whether or not it is an acceptable outcome. If it is acceptable, then the values are sent to the boiler, otherwise the data is sent back to the ML module for another prediction. Throughout this process, the data and outcomes are saved in order to use it as training data for the ML module.

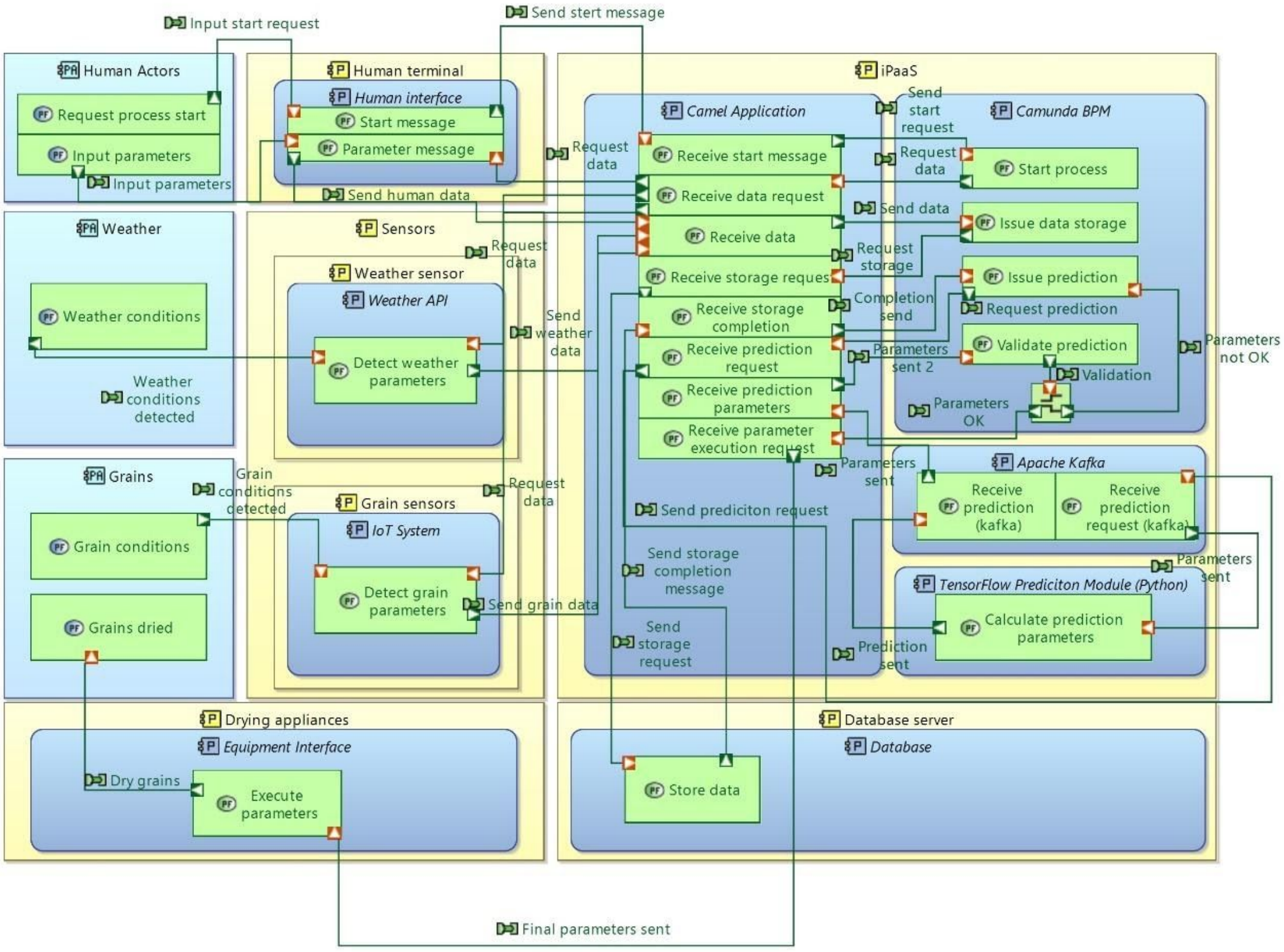

Fig. 2. Physical Architecture of the solution

The ML module itself uses historical data from previous operations in order to train its model for predictions based on desired output humidity, detected input humidity, and extraction weight. The dataset utilized contains the parameters from 153 individual drying cycles, with inputs from human experts in order to reduce overall energy consumption. The model is created after 5000 iterations and will define 16 rules, by default (adjustable), automatically according to the input data in addition to the rules that can be defined by an expert, utilizing 70% of the historical data as training data and the remaining 30% as test data to verify its prediction accuracy.

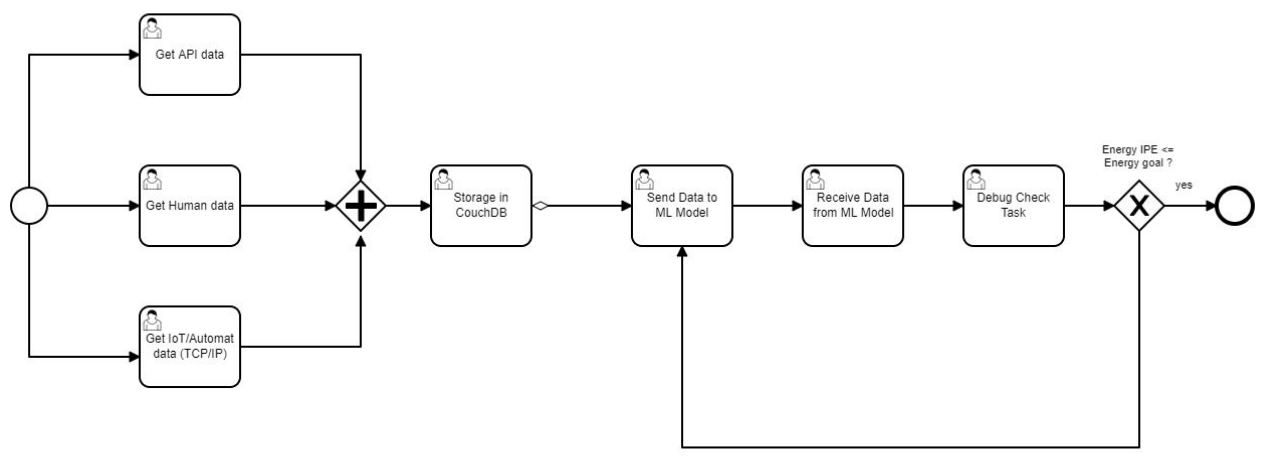

Fig. 3. BPM Diagram of the process

The predicted values are the extraction time, and the temperature settings for the dryer, which are also used in another step for the prediction of gas and energy consumption for the operation. The gas consumption value is the determining factor whether the predicted parameters are acceptable or not, as the main goals are not only to reduce the necessity of human interaction in the process, but also to guarantee that the overall energy consumption is acceptable or even reduced to optimal values. Due to time constraints, only the prediction of the extraction time was done by the proof-of-concept, using the value from the historical data as a comparison for the accuracy of the predicted value as well as the determining factor for the completion of a process cycle.

## V. EVALUATION AND RESULTS

In this context, we will focus on the data science dimension, more precisely, on the processing and knowledge resulting from the data. First, we will see an ANFIS based approach to predict the dryers setting parameters for the cereal drying process. Then, this prediction will be made using an ML model. Finally, the results will be presented with a comparison of the different approaches, according to key criteria for the company, demonstrating the advantages and disadvantages of each according to their needs and points of view.

## A. *ANFIS Solution*

The Adaptive Neuro Fuzzy Inference System approach was chosen allowing great flexibility with good results as well as a good understanding of predictions. In order to measure the loss, the Huber loss regression function is used because it is less sensitive to outliers in data than the squared error loss. It's also differentiable at 0. Huber loss approaches MSE when $\delta \sim 0$ and MAE when $\delta \sim \infty$ (large numbers) [24]. Regarding optimization, the Adam algorithm is used, which is an adaptive learning rate optimization algorithm that's been designed specifically for training deep neural networks. Adam can be seen as a combination of RMSprop and Stochastic Gradient Descent (SGD) with momentum. It uses squared gradients to scale the learning rate like RMSprop and it takes advantage of momentum by using moving average of the gradient instead of gradient itself like SGD with momentum. Today, Adam is definitely one of the best optimization algorithms for deep learning and its popularity is growing very fast [25].

This approach was implemented in Python and more precisely with the Tensorflow framework because it's an open-source framework which is complete and efficient.

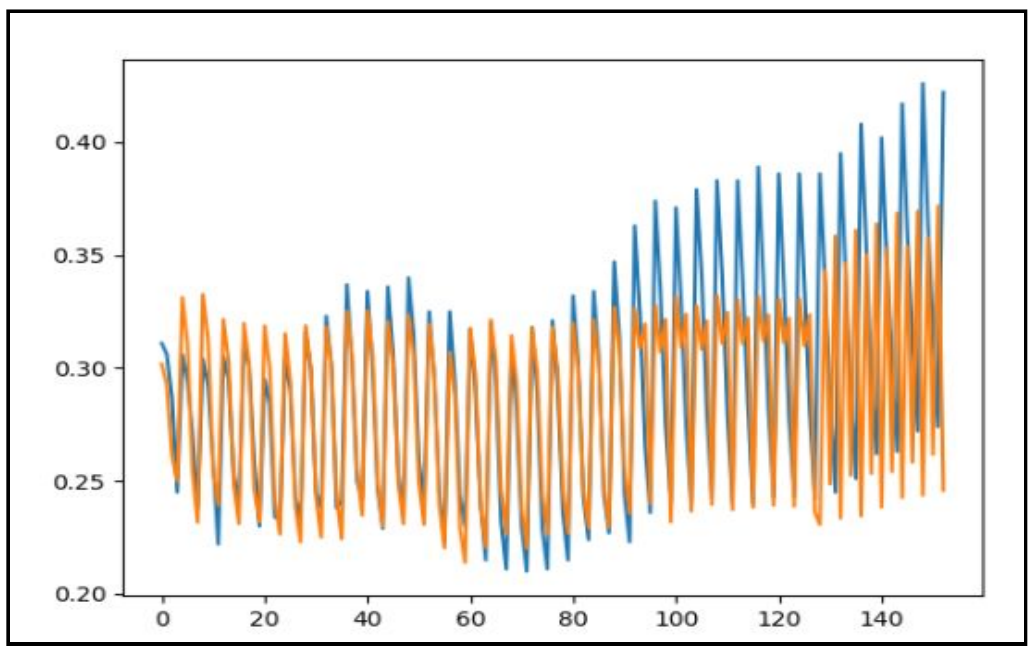

Fig. 4. Predicted (orange) vs Actual (Blue) Data

Figure 4 shows the predicted data compared to the actual data concerning the optimal settings to be defined in order to have good productivity while reducing energy consumption. As we can see, the model achieves good results with a small training sample.

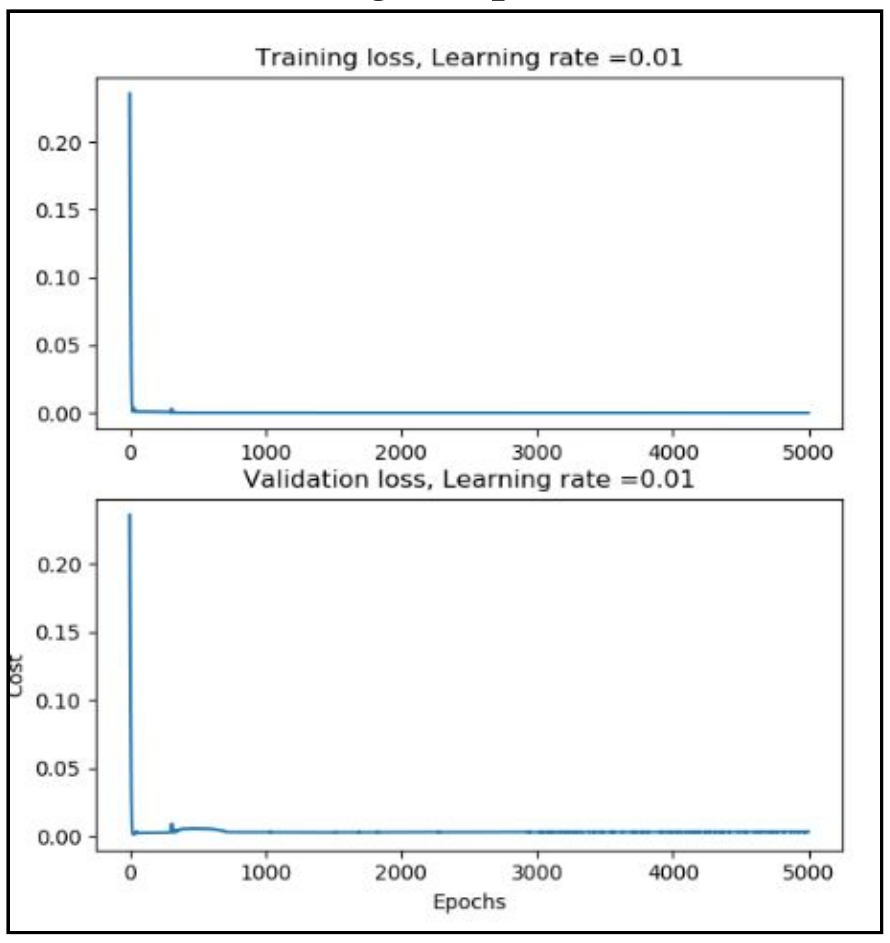

Fig 5. Training & Validation loss

Figure 5 shows the evolution of the relevance of the model for each generation after being optimized using the Adam algorithm. We can see that after 100 epochs, the model has practically reached its optimization limits. Here, we have defined 5000 epochs in order to be a little more precise. With a learning rate defined at 0.01, we can observe that the training and validation loss are close to 0 which represents a good result. In addition, the training time is rather fast since it amounts to only 12.7s.

## B. *ML Approach*

Regarding this approach, the exponential GPR model was chosen thanks to a study on several models. In order to implement it, the MATLAB tool was used, allowing a quick and complete analysis. The data used are obviously the same as for the previous approach in order to have a relevant comparison.

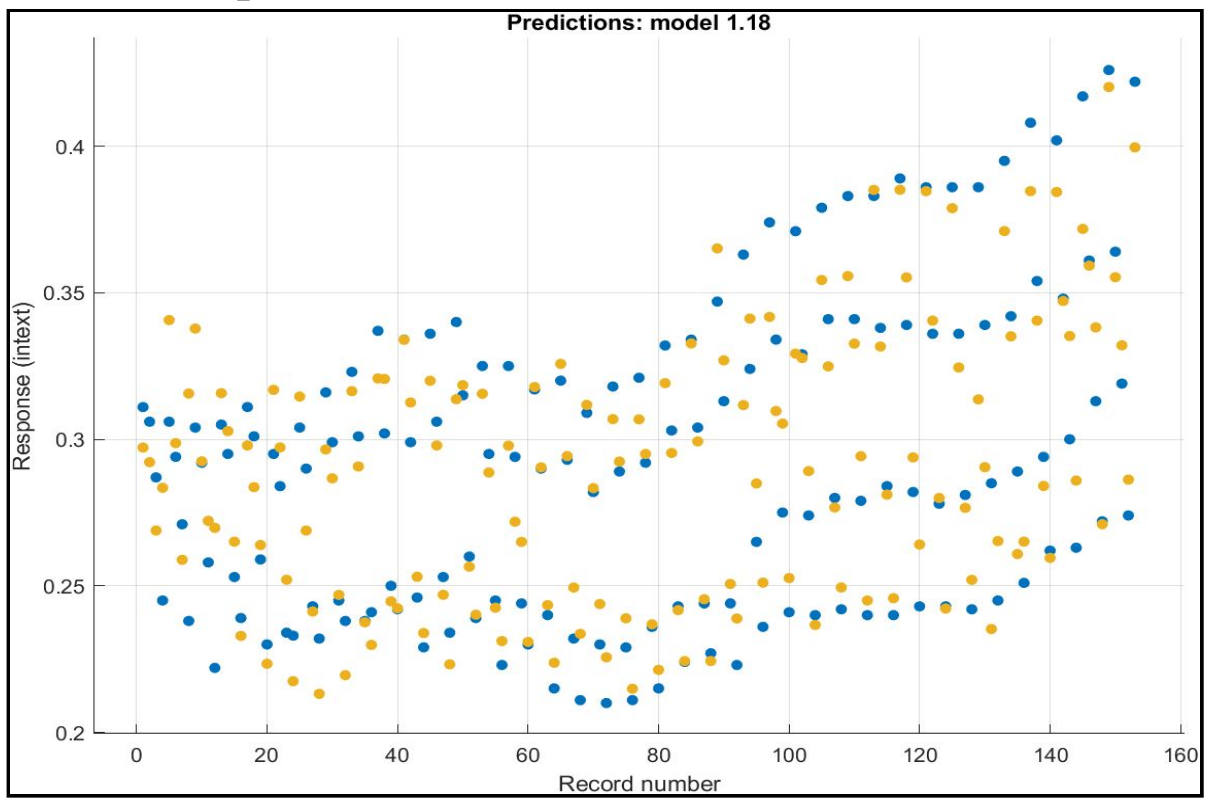

Fig. 6. Predicted (yellow) vs Actual (blue) Data

Figure 6 shows the predictions made in relation to the actual data, in particular : RMSE: 0.018865, R-Squared: 0.87, MSE: 0.0003559, MAE: 0.013531, Training time: 0.53s. The model presents very good results with a R-Squared amounting to 0.87 and RMSE, MSE and MAE close to 0, knowing that several optimizations are still possible, such as data filtering/cleaning, as well as increasing the number of individuals to be trained. In addition, the training time is really very fast with only 0.53s.

## C. *Approaches Comparison*

Both solutions have been used to meet an existential need to apply decisions based on real-time observations and reduce human intervention. Several performance and also industrial-oriented criteria were selected in order to compare these two approaches.

TABLE II. COMPARISON TABLE

| Criteria | ANFIS Solution | ML Solution |
|---|---|---|
| **Model Performance** | Satisfactory performance | Very satisfactory performance |
| **Explainability** | Auditable | Complex |
| **Model build speed** | Fast | Very Fast |
| **Model build complexity** | Simple | Simple |
| **Input data** | Not very sensitive to outliers | Very sensitive to outliers |
| **Flexibility** | Very flexible | Situational |

According to the above table based on previous experiences, we can say that both solutions are viable and have various advantages. The ML approach has a better performance and is faster to develop, however it provides little or no explicability and needs very clean data upstream. It's a really good approach to predict consumptions, for example, if you have a lot of reliable data. The ANFIS approach will therefore be more relevant if there is a need for auditability but also if there are specific rules to be considered where an expert can easily integrate them into the system. The latter is also much more flexible and can respond to a wider range of needs like the grain drying process where there are a lot of conditions and too much imprecise data.

### D. *Proof of concept*

The platform performed without issues, managing to automatically retrieve the necessary data, store it, and then send it for the prediction of the extraction time values. No stability issues were experienced during testing. Installation and adaptation required time and effort in order to understand the particularities of intercommunication between each component. While the solutions deployed have their intercommunication methods resolved in this prototype, adding new systems or components could take a few hours before being fully incorporated into the automated system.

## VI. Conclusions and perspectives

In this article, we have proposed a generic and open iPaaS architecture that is well-suited to satisfy integrability, interoperability and extensibility requirements as well as to allow the management of industry 4.0 real-time processes including well-adapted decision models.

We have illustrated how several decision models can be easily integrated and adapted in order to satisfy multiple applications and needs. For example, the fuzzy logic method would be perfect for monitoring or recommending industrial installations where many rules are already defined and known and where system interpretability is mandatory. On the other hand, the ML method is also very interesting for processes requiring high precision thanks to its very efficient models. In recent years, many applications have used fuzzy logic, optimization algorithms or neural networks in various fields. However, in most cases, a single decision model is not enough. If we take fuzzy logic, it can be said to provide the rough reasoning and constraint management which makes it flexible but has a low learning capacity, whereas on the other hand neural networks have this high learning capacity but have the incomprehensible "black box" syndrome. This is why we offer this architecture which allows smooth integration of the two models illustrated above but also offers the possibility of hosting several decision models, in parallel, which can be evaluated, adjusted and used simultaneously in order to help decision making. The interoperability of these models offers the possibility of capitalizing on their strengths while compensating for their weaknesses, which represents a major advantage for CPS who need high adaptability. The logical and physical architectures that were designed and implemented achieved a good foundation for an iPaaS, being robust and adaptable enough for reliable use in varied scenarios. In the future, the platform will be further developed in order to evaluate scalability and elasticity properties in multi-tenant scenarios.

## Acknowledgment

We thank MAÏSADOUR and UPPA for their financial and institutional support of the underlying research of this article.

## References

[1] N. Ebert, K. Weber, and S. Koruna, "Integration Platform as a Service," Bus Inf Syst Eng, vol. 59, no. 5, pp. 375–379, Oct. 2017, doi: 10.1007/s12599-017-0486-0.

[2] Cisco, "The Internet of Everything", Cisco, 2013. https://www.cisco.com/c/dam/en_us/about/business-insights/docs/ioe-value-at-stake-public-sector-analysis-faq.pdf (accessed Jul. 20, 2020).

[3] "The Maïsadour Cooperative Group in a nutshell." https://maisadour.eu/group/presentation/about-us/ (accessed Jul. 20, 2020).

[4] "Integration Platform as a Service (iPaas)." https://www.ibm.com/cloud/learn/ipaas (accessed Jul. 15, 2020).

[5] Gartner Inc., "iPaaS Integration Platform Software Reviews," Gartner. https://gartner.com/market/enterprise-integration-platform-as-a-service (accessed Jul. 08, 2020).

[6] Gartner Inc., "Gartner Reprint 'Magic Quadrant,'" Gartner Reprint, Apr. 23, 2019. https://www.gartner.com/doc/reprints?id=1-6KE0RUF&ct=190424&st=sb (accessed Jul. 08, 2020).

[7] Microsoft, "Azure - Enterprise Data Integration Patterns with Azure Service Bus." https://docs.microsoft.com/en-us/archive/msdn-magazine/2018/march/azure-enterprise-data-integration-patterns-with-azure-service-bus (accessed Jul. 15, 2020).

[8] Microsoft, "Messaging patterns - Cloud Design Patterns." https://docs.microsoft.com/en-us/azure/architecture/patterns/category/messaging (accessed Jul. 15, 2020).

[9] Microsoft, "Connector reference overview." https://docs.microsoft.com/en-us/connectors/connector-reference/ (accessed Jul. 15, 2020).

[10] "Integrations & Supported Applications," Boomi. https://boomi.com/platform/integration/applications/ (accessed Jul. 15, 2020).

[11] Boomi, "eBooks - eBook - Exploring Data Integration Patterns." https://resources.boomi.com/i/1159847-ebook-exploring-data-integration-patterns/3? (accessed Jul. 15, 2020).

[12] IBM, "IBM MQ." https://www.ibm.com/support/knowledgecenter/SS9H2Y_7.7.0/com.ibm.dp.doc/mq_introduction.html (accessed Jul. 15, 2020).

[13] IBM, "Connectors for App Connect | IBM." https://www.ibm.com/cloud/app-connect/connectors/ (accessed Jul. 15, 2020).

[14] "Message Exchange Patterns." https://docs.informatica.com/integration-cloud/cloud-application-integration/current-version/1----introduction/features/message-exchange-patterns.html (accessed Jul. 15, 2020).

[15] "All Cloud Connectors: Demos & Documentation | Informatica." https://www.informatica.com/products/cloud-integration/connectivity/connectors.html (accessed Jul. 15, 2020).

[16] "Workato connectors - PubSub by Workato | Workato Docs." https://docs.workato.com/connectors/pubsub.html#how-to-connect-to-the-pubsub-connector-on-workato (accessed Jul. 15, 2020).

[17] "Workato Integrations & Connectors," Workato. https://www.workato.com/integrations (accessed Jul. 15, 2020).

[18] "Capella MBSE Tool - Arcadia." https://www.eclipse.org/capella/arcadia.html (accessed Jul. 16, 2020).

[19] S. Cranton and J. Korab, Apache Camel Developer's Cookbook. Packt Publishing Ltd, 2013.

[20] "Workflow and Decision Automation Platform," Camunda BPM. https://camunda.com/ (accessed Jul. 07, 2020).

[21] "Apache Kafka," Apache Kafka. https://kafka.apache.org/intro (accessed Jul. 07, 2020).

[22] "TensorFlow," TensorFlow. https://www.tensorflow.org/ (accessed Jul. 07, 2020).

[23] J.-S. R. Jang, "ANFIS: adaptive-network-based fuzzy inference system," IEEE Transactions on Systems, Man, and Cybernetics, vol. 23, no. 3, pp. 665–685, May 1993, doi: 10.1109/21.256541.

[24] P. Grover, "5 Regression Loss Functions All Machine Learners Should Know," Medium, May 27, 2020. https://heartbeat.fritz.ai/5-regression-loss-functions-all-machine-learners-should-know-4fb140e9d4b0.

[25] V. Bushaev, "Adam — latest trends in deep learning optimization.," Medium, Oct. 24, 2018. https://towardsdatascience.com/adam-latest-trends-in-deep-learning-optimization-6be9a291375c.